# On Integrating Out Short-Distance Physics

## Vladimir Kalitvianski*

Grenoble, France

**Abstract**

I consider a special atomic scattering problem where the target atom has distinct "soft" and "hard" excitation modes. I demonstrate that in this problem the integration out of "short-distance" (or "high-energy") physics may occur automatically in the regular perturbative calculations, i.e., it may occur without any cut off and renormalization. Not only that, the soft inelastic processes happen already in the first Born approximation and the inclusive cross-sections become unavoidable from the very beginning. All that is possible because of correct physical and mathematical formulation of the problem. I propose to build QFT in a similar way.

**Keywords**

Cutoff, Renormalization, Reformulation, Effective Theory, Incomplete Theory, High-Energy Physics, Short-Distance Physics

## 1. Introduction

I would like to explain how short-distance (or high-energy) physics is "integrated out" in a reasonably constructed theory. Speaking roughly and briefly, there it is integrated out automatically. Neither cutoff nor renormalizations are necessary. On the other hand, the same theory may be formulated in such an awkward way that in order to obtain the same correct results some "renormalization" and summation of soft excitation contributions are obligatory starting from higher orders. As an example, I consider an old atomic scattering problem and solve it with the perturbation theory (Born series). The target atom (or ion) may be prepared in such a state that "soft" and "hard" atomic excitations are sufficiently distinct. Physically, the projectile may probe soft modes and at the same time it may be "ignorant" about the presence of hard ones. Mathematically it should be so too, but the latter depends on the theory formulation. Some physical theories (QFT) are formulated in such an awkward way that "brings forward" the inessential short-distance physics and prevents us from understanding how nature works. At the same time, the most probable events - soft excitations - are first missing in them. My atomic problem may help reformulate those theories in a better way since my problem can also be cast in a similar awkward formulation. The objective of this paper is to demonstrate the physically and mathematically reasonable approach.

Chapter 2 deals with the problem setup and phenomena to describe. It introduces the atomic form-factors and discusses their physics. In particular, it is shown that the short-distance physics may not influence the long-distance physics and it happens naturally.

Chapter 3 discusses another analogy with QED, namely, the soft excitation problem. In my approach there is no such a problem which is demonstrated with the "electronium" notion respecting the energy-momentum conservation law.

The awkward formulations with the forced soft contribution summation and constant renormalizations are discussed in Appendix.

## 2. Phenomena to Describe

Let us consider a two-electron Helium atom in the following state: one electron is in the "ground" state and the other one is in a high orbit. The total wave function of this system $\Psi(\mathbf{r}_{Nucl}, \mathbf{r}_{e_1}, \mathbf{r}_{e_2}, t)$ depending on the absolute coordinates $\mathbf{r}_{Nucl}$, $\mathbf{r}_{e_1}$ and $\mathbf{r}_{e_2}$, is conveniently presented as a product of a plane wave $\Phi(\mathbf{R}_A)e^{-iE_{P_A}t/\hbar}$, $\Phi(\mathbf{R}_A) = e^{i\mathbf{P}_A\mathbf{R}_A/\hbar}$ describing the atomic center of mass (subscript "A") and a wave function of the relative or internal collective motion of atomic constituents $\phi_n(\mathbf{r}_1, \mathbf{r}_2)e^{-iE_n t/\hbar}$, where $\mathbf{r}_a$ are the electron coordinates relative to the nucleus $\mathbf{r}_a = \mathbf{r}_{e_a} - \mathbf{r}_{Nucl}$, $a = 1, 2$ and $\mathbf{R}_A$ is the atomic center of mass coordinate: $M_A = (M_{Nucl} + 2m_e)$, $\mathbf{R}_A = \left[ M_{Nucl}\mathbf{r}_{Nucl} + m_e(\mathbf{r}_{e_1} + \mathbf{r}_{e_2}) \right]/M_A$, (see Fig. 1).

---

* Corresponding author
E-mail address: vladimir.kalitvianski@wanadoo.fr

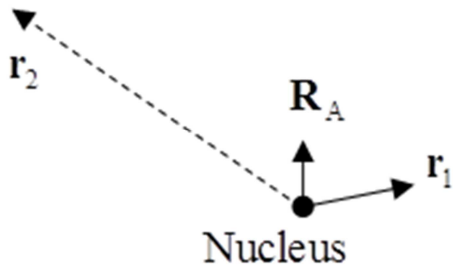

**Fig. 1.** Coordinates in question.

Normally, this wave function is still a complicated thing and the coordinates $\mathbf{r}_1$ and $\mathbf{r}_2$ are not separated (the interacting constituents are always in mixed states). What can be separated in $\phi_n(...)$ are normal (independent) modes of the collective motion (or "quasi-particles"). Normally, it is their properties (proper frequencies, for example) that are observed.

However, in case of one highly excited electron (a Rydberg state $n \gg 1$), the wave function of internal motion, for our numerical estimations and qualitative analysis, can be quite accurately approximated with a product of two hydrogen-like wave functions $\phi_n(\mathbf{r}_1, \mathbf{r}_2) \approx \psi_0(\mathbf{r}_1) \cdot \varphi_n(\mathbf{r}_2)$, where $\psi_0(\mathbf{r}_1)$ is a wave function of ion ($E1_0 \approx 2(E_H)_0$) since $Z_{Nucl} = 2$, and $\varphi_n(\mathbf{r}_2)$ is a wave function of Hydrogen in a highly excited state ($n \gg 1$, $E2_n \approx (E_H)_n$ since $(Z_{eff})_{He^+} = 1$, $E_n = E1_0 + E2_n$).

The system is at rest as a whole and serves as a target for a fast charged projectile (subscript "pr"). I want to consider large angle scattering, i.e., scattering from the atomic nucleus rather than from the atomic electrons. The projectile-nucleus interaction $V_{pr}(\mathbf{r}_{pr} - \mathbf{r}_{Nucl})$ is expressed via "collective" coordinates defined above thanks to the relationship $\mathbf{r}_{Nucl} = \mathbf{R}_A - m_e(\mathbf{r}_1 + \mathbf{r}_2)/M_A$.

I take a non-relativistic proton with $v \gg v_n$ as a projectile and I will consider such transferred momentum values $q = |\mathbf{q}|$ that are inefficient to excite the inner electron levels by "pushing" the nucleus. In other words, for the outer electron the proton is sufficiently fast to easily cause atomic transitions $\varphi_n \to \varphi_{n'}$ and to be reasonably treated by the perturbation theory in the first Born approximation, but for the inner electron the proton impact on the nucleus is such that it practically cannot cause the inner electron transitions, i.e., the main process for it is $\psi_0 \to \psi_0$. Below I will precise these conditions.

This two-electron atomic system will model a target with soft and hard target excitations, and the projectile is supposed to interact with one of its constituents – with the nucleus, via the Coulomb potential (i.e., no strong interactions are considered here). The scattering process can be schematically represented as follows: $pr + A \to pr' + A^*$, and the final states $pr'$ and $A^*$ are implied to be observable in some way, for example, with observing $\gamma$-decays of the excited target states $A^* \to A + \gamma$ and the Doppler shifts of $\gamma$ due to recoil.

### 2.1. Atomic Form-Factors

Now, let us look at the Born amplitude of scattering from such a target. The general formula for the cross-section is the following (all notations are taken from [1]):

$$d\sigma_{np}^{n'p'}(\mathbf{q}) = \frac{4m^2 e^4}{(\hbar q)^4} \frac{p'}{p} \cdot \left| Z_A \cdot f_n^{n'}(\mathbf{q}) - F_n^{n'}(\mathbf{q}) \right|^2 d\Omega, \quad (1)$$

$$F_n^{n'}(\mathbf{q}) = \int \phi_{n'}^*(\mathbf{r}_1, \mathbf{r}_2) \phi_n(\mathbf{r}_1, \mathbf{r}_2) \sum_a e^{-i\mathbf{q}\mathbf{r}_a} e^{i\frac{m_e}{M_A}\mathbf{q}\sum_b \mathbf{r}_b} d^3 r_1 d^3 r_2, \quad (2)$$

$$f_n^{n'}(\mathbf{q}) = \int \phi_{n'}^*(\mathbf{r}_1, \mathbf{r}_2) \phi_n(\mathbf{r}_1, \mathbf{r}_2) e^{i\frac{m_e}{M_A}\mathbf{q}\sum_a \mathbf{r}_a} d^3 r_1 d^3 r_2. \quad (3)$$

The usual atomic form-factor (2) describes scattering from atomic electrons (blue clouds in Fig. 2) and becomes relatively small for large scattering angles $\langle(\mathbf{q}\mathbf{r}_a)^2\rangle_n \gg 1$. It is so because, roughly speaking, the atomic electrons are light compared to the heavy projectile and they cannot cause large-angle scattering for a kinematic reason. I could consider scattering angles superior to those determined with the direct projectile-electron interactions $\theta \gg \frac{m_e}{M_{pr}} \frac{2v_0}{v}$, but for simplicity I exclude here the direct projectile-electron interactions $V_{pr}(\mathbf{r}_{pr} - \mathbf{r}_{e_a})$ in order not to involve $F_n^{n'}(\mathbf{q})$ in calculations at all (the electrons are "neutral" to our projectile, $F_n^{n'}(\mathbf{q}) = 0$). Then, for the projectile, there will be no nucleus charge "screening" due to atomic electrons nor atomic excitations due to direct projectile-electron interaction at any scattering angle (Fig. 3).

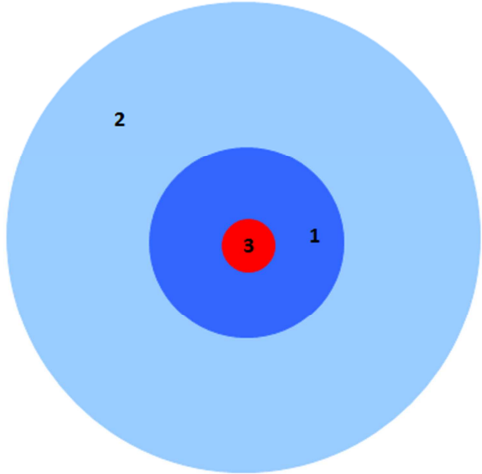

**Fig. 2.** Negative and positive "clouds" in our target schematically (scales and non-uniformity of $\psi_0$ and $\varphi_n$ are not respected): 1 – negative cloud of the first (inner) electron, 2 – that of the second one, 3 – positive cloud of the point-like nucleus bound in this system. This picture follows from formulas (1)-(3) in the elastic channel. The projectile in our consideration may only "see" the positive clouds: $d\sigma_{np}^{np}(\mathbf{q}) = d\sigma_{\text{Rutherford}} \cdot |f_n^n(\mathbf{q})|^2 d\Omega$.

Let us analyze the second atomic form-factor $f_n^n$ in the elastic channel $p' = p$ (the notion of a second atomic form-factor was first introduced in [1]). With our assumptions on the wave function $\phi_n(\mathbf{r}_1, \mathbf{r}_2)$, it can be easily calculated if the corresponding wave functions $\psi_0(\mathbf{r}_1)$ and $\varphi_n(\mathbf{r}_2)$ are injected in (3):

$$f_n^n(\mathbf{q}) \approx \int |\psi_0(\mathbf{r}_1)|^2 |\varphi_n(\mathbf{r}_2)|^2 e^{i\frac{m_e}{M_A}\mathbf{q}(\mathbf{r}_1+\mathbf{r}_2)} d^3r_1 d^3r_2. \quad (4)$$

It factorizes into two Hydrogen-like elastic form-factors:

$$f_n^n(\mathbf{q}) \approx f1_0^0(\mathbf{q}) \cdot f2_n^n(\mathbf{q})$$
$$= \int |\psi_0(\mathbf{r}_1)|^2 e^{i\frac{m_e}{M_A}\mathbf{q}\mathbf{r}_1} d^3r_1 \cdot \int |\varphi_n(\mathbf{r}_2)|^2 e^{i\frac{m_e}{M_A}\mathbf{q}\mathbf{r}_2} d^3r_2. \quad (5)$$

Form-factor $|f1_0^0(\mathbf{q})|$ describes quantum mechanical smearing of the nucleus charge due to nucleus coupling to the first atomic electron (a "positive charge cloud" 1 in Fig. 3). This form-factor may be close to unity – the charge smearing spot may look point-like to the projectile because of its small size $\propto (m_e/M_A) \cdot a_0/2$.

Form-factor $|f2_n^n(\mathbf{q})|$ describes quantum mechanical smearing of the nucleus charge ("positive charge cloud" 2 in Fig. 3) due to nucleus coupling to the second atomic electron. In our conditions $|f2_n^n(\mathbf{q})|$ is rather small because the corresponding smearing size $\propto (m_e/M_A) \cdot a_n$, $a_n \propto n^2$, $n \gg 1$ is much larger. In our problem setup the projectile "probes" these positive charge clouds and does not interact directly with the negative electrons (it does not "see" the blue clouds in Fig. 2).

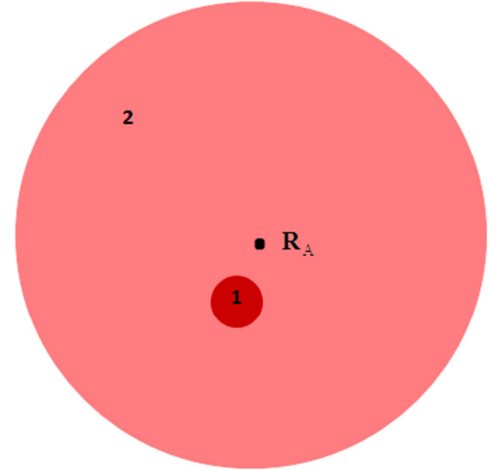

**Fig. 3.** Zoom of positive "clouds" in our target schematically (scales and non-uniformity of $\psi_0$ and $\varphi_n$ are not respected): 1 - positive cloud created with the point-like nucleus due to mutual motion with the first (inner) electron, 2 - that created mostly due to coupling to the second (outer) one. This picture is described with formulas (1), (4), (5) as long as the Born approximation is valid.

Thus, the projectile may "see" a big "positive charge cloud" (cloud 2 in Fig. 3) created with the motion of the atomic nucleus in its "high" orbit (i.e., with the motion of He$^+$ ion thanks to the second electron, but with full charge $Z_A = 2$ seen with the projectile), and at the same time it may not "see" the additional small positive cloud of the nucleus "rotating" also in the ground state of He$^+$ ion (cloud 1 in Fig. 3). Although cloud 2 is actually "drawn" with cloud 1, not with a point-like positive charge, the complicated short-distance structure (the small cloud within the large one) is integrated out in (5) and results in the elastic from-factor $|f1_0^0|$ tending to unity, as if its short-distance physics were absent and there only were a point-like nucleus "drawing" the second cloud: $d\sigma_{np}^{np}(\mathbf{q}) \approx d\sigma_{\text{Rutherford}} \cdot |f2_n^n(\mathbf{q})|^2 d\Omega$. We can choose such a proton energy $E_{\text{pr}}$ and such an excited state $\varphi_{n \gg 1}$, that $|f1_0^0|$ may be equal to unity even at the largest transferred momentum, i.e., at $\theta = \pi$.

### 2.1.1. Angle and Energy Dependencies

In order to see to what extent this is physically possible in our problem, let us analyze the "characteristic" angle $\theta 1_0$ for the inner electron state (formula (6) in [1]). (I remind that $q_{\text{elastic}}(\theta) = p \cdot 2\sin(\theta/2)$.) $\theta 1_0$ is an angle at which the inelastic processes become relatively essential – the probability of not exciting the target "inner" states is $|f1_0^0|^2$ and that of exciting any "inner" state is described with the factor $(1-|f1_0^0|^2)$ :

$$\theta 1_0 = 2\arcsin\left(\frac{2v_0}{2v}\cdot 5\right). \qquad (6)$$

Here, instead of $v_0$ stands $2v_0$ for the He$^+$ ion due to $Z_A = 2$, and factor 5 originates from the expression $(1+M_A/M_{pr})$. So, $\theta 1_0 = \pi$ for $v = 5v_0 = 2.5\cdot 2v_0$ ($E_{pr} = m_{pr}(5v_0)^2/2 \approx 0.63$ MeV, $E = m(5v_0)^2/2 \approx 0.5$ MeV). Fig. 4 shows just such a case: $f1_0^0(\mathbf{q})$ (the red line) together with the other form-factor $f2_3^3(\mathbf{q})$ (the blue line) – for a third excited state of the outer electron – in order to demonstrate a strong impact of $n$ on the smearing effect.

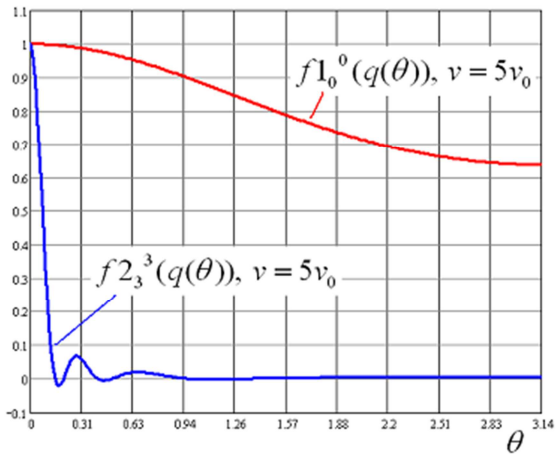

**Fig. 4.** Helium form-factors $f1_0^0$ and $f2_3^3$ at $v=5v_0$. $f1_0^0(q_{v=5v_0}(\pi))=0.64$.

We see that for scattering angles $\theta \ll \theta 1_0(v)$, i.e., where the most scattering events occurs, form-factor $|f1_0^0|$ becomes very close to unity – only elastic channel is open for the inner electron state and it results in a triviality as if there were no the inner electron with its states $\psi_n(\mathbf{r}_1)$ in our target. At the same time form-factor $|f2_n^n|$ may still be very small if $\theta \geq \theta 2_n \ll 1$. It describes a large and soft "positive charge cloud" in the elastic channel and for inelastic scattering $|f2_n^{n'}|$ describes the soft target excitations energetically accessible and efficient when pushing the heavy nucleus. Hence, one can observe no hard $\gamma$-quanta and plenty of soft ones in decays of A$^*$: $\varphi_{n'} \to \varphi_{n''} + \gamma$, where all $n''$, $n'$, and $n$ are implied to be much larger than 1.

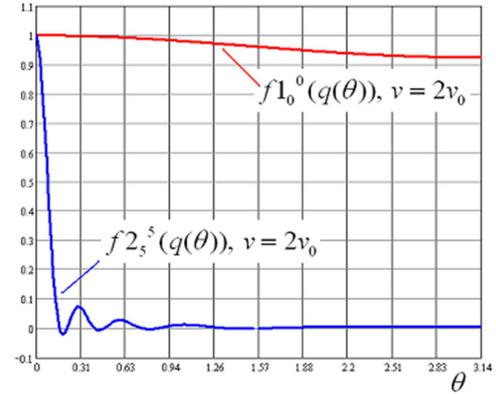

**Fig. 5.** Helium form-factors $f1_0^0$ and $f2_5^5$ at $v=2v_0$.

The inner electron level excitations due to hitting the nucleus can also be suppressed not only for $\theta \ll \theta 1_0(v)$, but also for large angles in case when the projectile velocities relatively small compared to the ground state electron velocity (Fig. 5).

(By the way, a light electron as a projectile does not see the additional small smearing even at $v = 10\cdot 2v_0$ because its energy is way insufficient and its de Broglie wavelength is too large for that. The incident electron should be rather relativistic to be able to probe such short-distance details [1].)

Let us note that for relatively small projectile velocities (namely, $v_n \ll v \leq 2v_0$) the first Born approximation may become somewhat inaccurate: the atomic nucleus may have enough time to make several small, but quick turns during interaction that leads to some minor "`polarization" of the "small positive spot" in Fig. 3 – the wave function of He$^+$ ion $\psi_0$ is slightly modified during "quasi-adiabatic" interaction, and this effect influences numerically the exact elastic cross section. The higher-order perturbative corrections of the Born series take care of this effect, but the short-distance physics will still not intervene in a harmful way in our calculations since it is already "out of reach". Instead of simply dropping out (i.e., producing a unity factor at the (Rutherford) cross section (1)), it will be taken into account ("integrated out") more precisely, if necessary. (The corresponding scattering physics is comprehensible in the opposite – Born-Oppenheimer approximation and simply "integrating it out" a la Wilson needs more careful justification in order to be convincing.)

### 2.1.2. Insensitivity to Short-Distance Physics

Hence, whatever the true internal structure is (the true high-energy physics, the true high-energy target excitations), the projectile in our "two-electron" theory cannot factually probe it when it effectively lacks energy for good resolution. The soft excitations are accessible and the hard ones are not. It is comprehensible physically and is rather natural – the projectile, as a long wave, only "sees" large things. Small details are somehow "averaged" or "integrated out". (Here I am excluding on purpose the fine and other kinds of level splitting from the hard spectrum of the target; otherwise transitions between them might become accessible!)

In our calculation this "integrating out" (factually, "taking into account") the short-distance physics occurs automatically rather than "manually". We do not introduce a cut-off and do not discard ("absorb") the harmful corrections in order to obtain something physical. We do not have harmful corrections at all. This convinces me in a possibility of constructing a physically reasonable QFT where no cut-off and discarding (renormalization) are necessary (see Appendix, especially A.5., for technical details).

The first Born approximation (3) in the elastic channel gives a "photo" of the atomic positive charge distribution, as if the atom was internally unperturbed during scattering; a photo with a certain resolution, though. Although the scattering amplitude depends on $\mathbf{q}$ in a more complicated way than just a Fourier transform of the Coulomb potential $\propto e^2/q^2$, I do not assign the additional $\mathbf{q}$-dependence to the nucleus charge $Ze$ or to something else. I.e., I do not introduce running constants. I do not say that in terms of the effective elastic potential [1] I have some charge "anti-screening" like in QCD. I say that the effective potential behavior at short distances (Fig. 1 and Fig. 4 in [1]) is a typical effect of quantum mechanical smearing in a compound system.

### 2.1.3. Inelastic and Inclusive Cross Sections

Inelastic processes $n' \neq n$ produce possible final target states different from the initial one (different could 2 configurations in Fig. 3).

The fully inclusive cross section (i.e., the sum of the elastic and all inelastic ones) reduces to a great extent to a Rutherford scattering formula for a free and still point-like target nucleus – no "clouds" at all: $V_{\text{pr}}(\mathbf{r}_{\text{pr}} - \mathbf{r}_{\text{Nucl}}) \to V_{\text{eff}} \approx V_{\text{pr}}(\mathbf{r}_{\text{pr}} - \mathbf{R}_A) \propto 1/|\mathbf{r}|$, see formula (9) in [1]. (Here I imply the scattering angles $\theta 2_n \leq \theta \ll \theta 1_0$ and summing up on $\varphi_{n'}$ solely. Otherwise ($\theta \geq \theta 1_0$) the "cloud" $|f1_0^0(\mathbf{q})|^2 < 1$ and inelastic amplitudes $\propto f1_0^{n'}(\mathbf{q})$ will intervene too.)

The inclusive picture is *another kind of averaging* – over the whole variety of events, averaging often encountered in experiments and resulting in a deceptive simplification. One has to keep this in mind because usually it is not mentioned while speaking of short-distance physics, as if there were no difference between elastic, inelastic, and inclusive pictures! It is crucial to distinguish them in the correct physical description.

Increasing the projectile energy (decreasing its de Broglie wavelength), increasing the scattering angles and resolution at experiment help reveal the short-distance physics in more detail. Doing so, we may discover high-energy excitations inaccessible at lower energies/angles. As well, we may learn that our knowledge (for example, about point-likeness of the core) was not really precise, "microscopic", but inclusive ($V_{\text{eff}}(\mathbf{r})$ is not "microscopic" and exhaustive). And, of course, the symmetry of the high-energy physics may well be different from that of the low-energy physics. One can understand the latter property as a "symmetry breaking" at high/low energies.

### 2.2. Absence of Mathematical and Physical Difficulties

Above we did not encounter any mathematical difficulties. It was a banal calculation, as it should be in physics. We may therefore say that our theory is physically reasonable.

What does make our theory physically reasonable? Its correct formulation. The permanent interactions of the atomic constituents is taken into account exactly, both via their wave function and via the relationships between their absolute and the relative (or collective) coordinates, namely, $\mathbf{r}_{\text{Nucl}}$ involved in $V_{\text{pr}}(\mathbf{r}_{\text{pr}} - \mathbf{r}_{\text{Nucl}})$ was expressed via $\mathbf{R}_A$ and $\mathbf{r}_a$. The rest was a perturbation theory in this or that approximation. For scattering processes it calculates the occupation number evolutions – the transition probabilities between different target and projectile states. Even in the first Born approximation all possible target excitations are present in a non-trivial and reasonable way – via form-factors. It is an ideal situation in the scattering physics description. I say so because for the same problem there may be awkward "descriptions" too – with its weird "physics" (see Appendix).

Now, let us imagine for instance that "there is nothing in the world but out target and the projectile", and our "two-electron" theory above is then a "Theory of Everything" (or a true "underlying theory") unknown to us so far. Low-energy experiments outlined above would not reveal the "core" structure, but would present it as a point-like nucleus smeared only due to the second electron. Such experiments would then be well described with a simpler, "one-electron" theory, a theory of a hydrogen-like atom with $\varphi_n(\mathbf{r}_2)$ and $M_A$. The presence of the first (inner) electron would not be necessary in such a theory: the latter would work fine and without difficulties – it would reproduce low-energy target excitations if we could guess the simplified theory right.

May we call the "one-electron" theory an effective one? Maybe. I prefer the term "incomplete" – it does not include and predict all target excitations existing in our simplified "nature", but it has no mathematical problems (catastrophes) as a model even outside its domain of validity, i.e. for $\theta \geq \theta l_0$. The Born series terms all are finite and the projectile energy $E_{pr}$ (or a characteristic transferred momentum $|\mathbf{q}|$) is not a "scale" in our theory in a Wilsonian sense. This is (A.4) – a reformulated, physically meaningful theory with respect to awkward formulations presented in Appendices A.2., A.4., and A.5.

Thus, the absence of the true physics of short distances in the "one-electron" theory does not make it ill-defined or fail mathematically. And this is so because the one-electron theory is also constructed correctly – what is know to be coupled permanently and determines the soft spectrum is already taken into account in it via the wave function $\varphi_n(\mathbf{r}_2)$ and via the coordinate relationships. That is why when people say that a given theory has mathematical problems "because not everything in it is taken into account", I remain skeptic. I think the problem is in its erroneous formulation. It is a problem of formulation or modeling (see, for example, unnecessary and harmful "electron self-induction effect" discussed in [2] and an equation coupling error discussed in [3]). And I do not believe that when "everything else" is taken into account, the difficulties will disappear automatically. Especially if "new physics" is taken into account in the same way – erroneously. Instead of excuses, we need more correct formulations of incomplete theories on each level of our knowledge. (And there may be a plenty of such alternative formulations, as a matter of fact.)

## 3. Analogy with QED

### 3.1. Analogy of Inelastic Processes

Now, let us turn to QED and consider a charge-one state in it, normally associated with one electron, at rest. According to QED equations, "everything is permanently coupled with everything else", in particular, even one-electron (i.e., charge-1) state, as a target, contains possibilities of exciting high-energy states like creating hard photons and electron-positron pairs. It is certainly so in experiments, but the standard QED suffers from calculation difficulties (catastrophes) of obtaining them in a natural way because of its awkward formulation, in particular, because of too bad initial approximations (see Appendices for explanation). A great deal of QED calculations consists in correcting its initial wrongness. That is why "guessing right equations" is still an important physical and mathematical task.

### 3.2. Electronium and All That

My electronium model [1] is an attempt to take into account a low-energy QED physics, like in the "one-electron" incomplete atomic model mentioned briefly above. The non-relativistic electronium model $\Psi = e^{i\mathbf{PR}/\hbar} \phi_n(Q_{\mathbf{k}_1,\lambda_1},...) e^{-iE_n t/\hbar}$ does not include all possible QED excitations but soft photons; however, and this is important, it works fine in a low-energy region. Colliding two electroniums produces soft excitations (radiation) immediately, in the first Born approximation. It looks like colliding two complex atoms – in the final state one naturally obtains excited atoms. By the way, in my opinion, the electromagnetic field oscillators are those normal modes of the collective motions whose variables in the corresponding $\phi_n(Q_{\mathbf{k}_1,\lambda_1},...)$ of electronium are separated: $\phi_n(Q_{\mathbf{k}_1,\lambda_1},...) = \prod_{\mathbf{k},\lambda} \chi_{\mathbf{k},\lambda}(Q_{\mathbf{k},\lambda})$ (see (16) in [1]).

There is no background for the infrared problem there because the soft modes are taken into account "exactly" rather than "perturbatively". Perturbative treatment of soft modes in QED gives a divergent series due to "strongness" of soft mode contributions into the calculated probabilities [4]:

---

§ 30. Emission of Photons of Long Wavelength

30.1. *The "infrared catastrophe"*

In the preceding section we have seen that the probability of the photon emission in the low energy range is inversely proportional to the frequency $dw \propto (d\omega/\omega)$, while the total radiation probability diverges logarithmically as $\omega \to 0$.

This divergence in the region of low photon energies is referred to as the "infrared catastrophe".

It is brought about by unjustified application of ordinary perturbation theory, based on the expansion of the scattering matrix into a series in powers of $e$, to those processes in which long wavelength photons participate. Indeed, it can be easily shown that if the probability $w_1$ for emission of one long wavelength photon is proportional to $e^2 \ln(\varepsilon/\omega)$ where $\varepsilon$ is an energy of the order of magnitude of the electron energy, then the probability of emission of two photons will be proportional to $(e^2 \ln(\varepsilon/\omega))^2$. Therefore, the ratio of probabilities is given in order magnitude by

$$\xi \equiv \frac{w_1}{w_2} \propto e^2 \ln\frac{\varepsilon}{\omega} \text{ where } \omega \to 0. \quad (30.1)$$

It is this ratio, and not the quantity $e^2$ as we have been assuming until now, which provides the perturbation theory expansion parameter applicable to the processes of interaction between the electron and long wavelength photons. Since when $\omega \to 0$, $\xi$ is not small compared to unity, perturbation theory is, strictly speaking, inapplicable to such cases.

---

**Fig. 6.** Extraction from [4].

As electronium is constructed by analogy with atom, here there is a direct analogy with our atomic target which is easy to note with expanding our second form-factors $f_n^{n'}(\mathbf{q})$ in powers of "small coupling constant" $m_e/M_A$ in the exponential (3), for example: $f_n^n(\mathbf{q}) \approx 1 - \frac{m_e^2}{M_A^2}\langle(\mathbf{qr}_a)^2\rangle_n$. For the first electron (i.e., for the hard excitations) the term $\frac{m_e^2}{M_A^2}\langle(\mathbf{qr}_1)^2\rangle_0$ may be small (see Fig. 5) whilst for the second one $\frac{m_e^2}{M_A^2}\langle(\mathbf{qr}_2)^2\rangle_n$ may be rather large and it diverges in the soft limit $n \to \infty$ anyway. Here, like $\alpha$ in QED, the small dimensionless "coupling constant" $m_e/M_A$ never comes alone, but with another dimensionless factor – a function of the problem parameters, so such perturbative corrections may take any value. In QED the hard and soft photon modes, i.e., "small" and "big" corrections, are both treated perturbatively because the corresponding electron-field interaction is factually written separately – in the so called "mixed variables" [5] and the corresponding QED series are similar to expansions of our form-factors $f_n^{n'}$ in powers of $m_e/M_A$ (see Appendices A.1. and A.2.).

How could one complete my electronium model? One could add all QED excitations in a similar way – as a product of the other possible "normal modes" to the soft photon wave function and express the constituent electron coordinates via the center of mass and relative motion coordinates, like in the non-relativistic electronium or in atom. Such a completion would work as fine as my actual (primitive) electronium model, but it would produce the whole spectrum of possible QED excitations in a natural way. Such a reformulated QED model would be free from mathematical and conceptual difficulties *by construction*. Yes, it would be still an "incomplete" QFT, but no references to the absence of the other particles (excitations) existing in Nature would be necessary. No artificial cut-off with integrating out "fast modes" [6] and introducing running constants [7] would be necessary in order to get rid of initial wrongness, as it is carried out today in the frame of Wilsonian RG exercise in QFT.

## 4. Conclusions

In a "complete" reformulated QFT (or "Theory of Everything") the "non-accessible" at a given energy $E$ excitations would not contribute (with some reservations). Roughly speaking, they would be integrated out (taken into account) automatically, like in my "two-electron" target model given above, reducing naturally to a unity factor or so. But this property of "insensitivity to short-distance physics" does not exclusively belong to the "complete" reformulated QFT. "Incomplete" theories can also be formulated in such a way that this property will hold. It means the short-distance physics, present in such an "incomplete theory" and different from reality, cannot be and will not be harmful for calculations technically, as it was eloquently demonstrated in this article. When the time arrives, the new high-energy excitations could be taken into account in a natural way, roughly speaking, as a transition from a "one-electron" to "two-electron" target model above. I propose to think over this way of constructing QFT. I feel it is a promising direction of building physical theories.

## Appendix

### A.1. Typical (Collective) Variables

Formulation in terms of mixed variables consists in using an "individual" coordinate of one of constituent particle and relative coordinates for the other particles. To explain the corresponding physics and techniques, let us consider a simple two-particle system as a target, a Hydrogen atom, for example. The target Hamiltonian can be written via the individual and "collective" coordinates (no mixed variables so far):

$$\hat{H}_H = -\frac{\hbar^2}{2M_{Nucl}}\frac{\partial^2}{\partial \mathbf{r}_{Nucl}^2} - \frac{\hbar^2}{2m_e}\frac{\partial^2}{\partial \mathbf{r}_{e_1}^2} + V_A(\mathbf{r}_{e_1} - \mathbf{r}_{Nucl}), \quad (A.1)$$

$$\hat{H}_H = -\frac{\hbar^2}{2M_A}\frac{\partial^2}{\partial \mathbf{R}_A^2} + \left[-\frac{\hbar^2}{2\mu}\frac{\partial^2}{\partial \mathbf{r}_1^2} + V_A(\mathbf{r}_1)\right]. \quad (A.2)$$

In the latter case (A.2) the coordinates $\mathbf{R}_A$ and $\mathbf{r}_1$ are separated and the Hamiltonian provides the spectrum of the target states as a product $|\mathbf{P}_A, n\rangle = |\mathbf{P}_A\rangle|n\rangle$.

When we add a projectile interacting with the nucleus:

$$-\frac{\hbar^2}{2M_{pr}}\frac{\partial^2}{\partial \mathbf{r}_{pr}^2} + V_{pr}(\mathbf{r}_{pr} - \mathbf{r}_{Nucl}), \quad (A.3)$$

the total Hamiltonian may read as follows (here we introduce "scattering" variables: $\mathbf{R}_{CI} = (M_{pr}\mathbf{r}_{pr} + M_A\mathbf{R}_A)/M_{tot}$, $\mathbf{r} = \mathbf{r}_{pr} - \mathbf{R}_A$, $M_{tot} = M_{pr} + M_A$, and $m = M_{pr}M_A/M_{tot}$):

$$\hat{H}_{tot} = -\frac{\hbar^2}{2M_{tot}}\frac{\partial^2}{\partial \mathbf{R}_{CI}^2} + \left[-\frac{\hbar^2}{2\mu}\frac{\partial^2}{\partial \mathbf{r}_1^2} + V_A(\mathbf{r}_1)\right] + \left[-\frac{\hbar^2}{2m}\frac{\partial^2}{\partial \mathbf{r}^2} + V_{pr}\left(\mathbf{r} + \frac{m_e}{M_A}\mathbf{r}_1\right)\right]. \quad (A.4)$$

The first term in (A.4) describes a typical free motion of the center of inertia of the total system (projectile + atom), and it provides the total energy and momentum conservation during scattering (the scattering potential $V_{pr}$ does not depend on $\mathbf{R}_{CI}$ at all).

The first square bracket in (A.4) is a typical textbook Hamiltonian for the Hydrogen eigenfunctions $\psi_n(\mathbf{r}_1)$ and eigenvalues $E_n$. Here $\mu = m_e M_{Nucl}/(m_e + M_{Nucl})$.

The last square bracket in (A.4) is a typical textbook Hamiltonian describing the scattering problem in the global CI coordinates. Without $V_{pr}$ all variables in (A.4) are separated. This fact helps build the perturbation theory in powers of $V_{pr}$. The only difference between our expression (A.4) and the textbook one is in the presence of a "small" term $\frac{m_e}{M_A}\mathbf{r}_1$ in the interaction potential argument. I did not neglect it because this term is necessary for the projectile to act on the nucleus: $\mathbf{r}_{pr} - \mathbf{r}_{Nucl} = \mathbf{r} + \frac{m_e}{M_A}\mathbf{r}_1$. Without this "small" term the projectile transfers its momentum to the atomic center of mass: $V(\mathbf{r}_{pr} - \mathbf{r}_{Nucl}) \to V(\mathbf{r} = \mathbf{r}_{pr} - \mathbf{R}_A)$, and thus it cannot cause atomic excitations no matter how big the transferred momentum is – the atom is accelerated as a whole. In other words, without this "small term" only elastic scattering from a point-like atomic center of mass occurs: $f_n^{n'}(\mathbf{q}) = \delta_{nn'}$ which is unphysical for any compound target.

In the main text such "small terms" are taken into account "exactly" (as long as the first Born approximation applies) which gives non trivial and physically correct atomic form-factors (2), (3).

### A.2. Mixed Variables I – Formulation with the Infra-Red Problem

However, there may be a formulation where this "small term" is forced to be taken into account perturbatively so that the first Born approximation becomes somewhat unphysical, like in QED. I am not speaking here of literally expanding the interaction potential $V_{pr}$ in powers of $m_e/M_A$ in the Hamiltonian (A.4). I am speaking of a formulation where this term stands in the Hamiltonian as an additional operator. In order to explain this point, let us introduce mixed variables, say, the individual nucleus coordinate and the relative electron-nucleus coordinate:

$$\mathbf{R}' = \mathbf{r}_{Nucl},\; \mathbf{r}_1' = \mathbf{r}_{e_1} - \mathbf{r}_{Nucl},\; \frac{\partial}{\partial \mathbf{r}_{Nucl}} = \frac{\partial}{\partial \mathbf{R}'} - \frac{\partial}{\partial \mathbf{r}_1'},\; \frac{\partial}{\partial \mathbf{r}_{e_1}} = \frac{\partial}{\partial \mathbf{r}_1'}. \quad (A.5)$$

The Hydrogen Hamiltonian may be rewritten in the following way (see some other ways in the next subsection):

$$\hat{H}_H = -\frac{\hbar^2}{2M_A}\frac{\partial^2}{\partial \mathbf{R}'^2} + \left[-\frac{\hbar^2}{2\mu}\frac{\partial^2}{\partial \mathbf{r}_1'^2} + V_A(\mathbf{r}_1')\right] \\ -\frac{\hbar^2}{2M_{Nucl}}\left(\frac{m_e}{M_A}\frac{\partial^2}{\partial \mathbf{R}'^2} - 2\frac{\partial}{\partial \mathbf{R}'}\frac{\partial}{\partial \mathbf{r}_1'}\right). \quad (A.6)$$

The first three terms (the first line) have the same functional form as the Hydrogen Hamiltonian in the "collective" coordinates (A.2) and they give solutions of the same analytical structure, namely, a product of a plane wave $e^{i(\mathbf{PR}' - E_p t)/\hbar}$ and the Hydrogen wave function $\psi_n(\mathbf{r}_1')e^{-iE_n t/\hbar}$, with $\mathbf{P}$ being the total momentum of the atom (target momentum).

The presence of the round-bracket term in (A.6) $-\frac{\hbar^2}{2M_{Nucl}}\left(\frac{m_e}{M_A}\frac{\partial^2}{\partial \mathbf{R}'^2} - 2\frac{\partial}{\partial \mathbf{R}'}\frac{\partial}{\partial \mathbf{r}_1'}\right) \equiv \delta\hat{H}$ indicates that the mixed variables, despite being independent, are not separated. In particular, this term takes into account the difference between a free atomic center of mass motion $e^{i(\mathbf{PR}_A - E_p t)/\hbar}$ and the inexact plane wave describing a free motion of the nucleus: $\mathbf{R}_A = \mathbf{R}' + \frac{m_e}{M_A}\mathbf{r}_1' = \mathbf{r}_{Nucl} + \frac{m_e}{M_A}\mathbf{r}_1$, i.e., it takes into account the missing factor $\exp\left(i\frac{m_e}{M_A}\mathbf{P}\cdot\mathbf{r}_1/\hbar\right)$ containing the dimensionless "coupling constant" $m_e/M_A$ and making the nucleus "rotate" in the atom instead of moving uniformly.

Note, in the plane wave $\Phi(\mathbf{R}_A) = e^{i\mathbf{PR}_A/\hbar}$ the individual coordinates of the atomic constituents come with the corresponding weights: $\mathbf{R}_A = \frac{M_{Nucl}}{M_A}\mathbf{r}_{Nucl} + \frac{m_e}{M_A}\mathbf{r}_{e_1}$; however expressed via the mixed variables the plane wave looks differently: $\mathbf{R}_A = \mathbf{r}_{Nucl} + \frac{m_e}{M_A}\mathbf{r}_1$ since a part of $\mathbf{r}_{Nucl}$ is already contained in $\mathbf{r}_1$. Therefore, the wave $\psi(\mathbf{r}_1)$ becomes a wave of the relative motion in a compound system rather than a wave "independent" of the nucleus.

Résumé: although legitimate exactly, perturbatively (A.6) is already a bad formulation as it generates corrections to the zeroth-order solutions. The latter describe the motion of a free nucleus $\Phi^{(0)}(\mathbf{R}' = \mathbf{r}_{Nucl}) = e^{i\mathbf{Pr}_{Nucl}/\hbar}$ with a "mass" $M_A$ (!) decoupled from the target "internal" degrees of freedom. Hence, the perturbative corrections due to operator $\delta\hat{H}$ modify not only numerical values, but also the physical meaning of the target solutions – the exact plane wave describes a free motion of the center of mass of permanently interacting system and the wave $\psi(\mathbf{r}_1')$ with its observable frequencies describes in fact a relative motion in this compound system rather than something "independent" of the nucleus. This is a nontrivial lesson of the "coupling physics" in this formulation, and I believe that such an understanding is needed in QFT too.

Now, when we add a projectile (A.3) to (A.6), the total Hamiltonian can be cast in the form:

$$\mathbf{R}_{CI}' = (M_{pr}\mathbf{r}_{pr} + M_A\mathbf{R}')/M_{tot}, \quad \mathbf{r}' = \mathbf{r}_{pr} - \mathbf{R}', \quad (A.7)$$

$$\hat{H}_{tot} = -\frac{\hbar^2}{2M_{tot}}\frac{\partial^2}{\partial \mathbf{R}_{CI}'^2}$$
$$+ \left[-\frac{\hbar^2}{2\mu}\frac{\partial^2}{\partial \mathbf{r}_1'^2} + V_A(\mathbf{r}_1')\right] + \left[-\frac{\hbar^2}{2m}\frac{\partial^2}{\partial \mathbf{r}'^2} + V_{pr}(\mathbf{r}')\right] \quad (A.8)$$
$$+ \delta\hat{H}.$$

The first two lines in (A.8) coincide formally with the total Hamiltonian (A.4) where the "small term" $\frac{m_e}{M_A}\mathbf{r}_1$ in $V_{pr}$ is absent. It is the third line $\delta\hat{H}$ who is present in the total Hamiltonian instead. This term is out of the potential $V_{pr}$ argument, i.e., it is an additional operator in the Schrödinger equation – it "enlarges" "free" Hamiltonians (like any gauge interaction does). It means that if we build the Born series for the scattering problem, the interaction potential $V_{pr}(\mathbf{r}')$ itself will give unphysical elastic amplitude $\propto f_n^{n'}(\mathbf{q}) = \delta_{nn'}$ in the first Born approximation no matter how large the transferred momentum is. It also means that it is the term $\delta\hat{H}$ who will give perturbative corrections to $\delta_{nn'}$ which were briefly and partially outlined in the paragraph just below Figure 6. The calculation, however, will now be more complicated in comparison with a simple Taylor expansion of the exponential in (3) since, considered perturbatively together with $V_{pr}(\mathbf{r}')$ (or with an external potential $V_{ext}(\mathbf{r}')$), this term contributions will appear in higher orders and together with higher powers of $V_{pr}(\mathbf{r}')$, so one will need to rearrange the calculated terms in order, for example, to group some of them into $f_n^n(\mathbf{q}) \approx 1 - \frac{m_e^2}{M_A^2}\langle(\mathbf{qr}_1)^2\rangle_n$.

In practical calculations within the formulation (A.8), the term $\delta\hat{H}$ should also be expressed via the new (scattering) variables (A.7):

$$\frac{\partial}{\partial \mathbf{r}_{pr}} = \frac{M_{pr}}{M_{tot}}\frac{\partial}{\partial \mathbf{R}_{CI}'} + \frac{\partial}{\partial \mathbf{r}'},$$
$$\frac{\partial^2}{\partial \mathbf{r}_{pr}^2} = \frac{M_{pr}^2}{M_{tot}^2}\frac{\partial^2}{\partial \mathbf{R}_{CI}'^2} + \frac{\partial^2}{\partial \mathbf{r}'^2} + 2\frac{M_{pr}}{M_{tot}}\frac{\partial}{\partial \mathbf{R}_{CI}'}\frac{\partial}{\partial \mathbf{r}'} \quad (A.9)$$

$$\frac{\partial}{\partial \mathbf{R}'} = \frac{M_A}{M_{tot}}\frac{\partial}{\partial \mathbf{R}_{CI}'} - \frac{\partial}{\partial \mathbf{r}'},$$
$$\frac{\partial^2}{\partial \mathbf{R}'^2} = \frac{M_A^2}{M_{tot}^2}\frac{\partial^2}{\partial \mathbf{R}_{CI}'^2} + \frac{\partial^2}{\partial \mathbf{r}'^2} - 2\frac{M_A}{M_{tot}}\frac{\partial}{\partial \mathbf{R}_{CI}'}\frac{\partial}{\partial \mathbf{r}'}, \quad (A.10)$$

$$\delta\hat{H} \propto \frac{m_e}{M_A}\left(\frac{M_A^2}{M_{tot}^2}\frac{\partial^2}{\partial \mathbf{R}_{CI}'^2} + \frac{\partial^2}{\partial \mathbf{r}'^2} - 2\frac{M_A}{M_{tot}}\frac{\partial}{\partial \mathbf{R}_{CI}'}\frac{\partial}{\partial \mathbf{r}'}\right)$$
$$- 2\left(\frac{M_A}{M_{tot}}\frac{\partial}{\partial \mathbf{R}_{CI}'} - \frac{\partial}{\partial \mathbf{r}'}\right)\frac{\partial}{\partial \mathbf{r}_1'}. \quad (A.11)$$

Now one can use formula (43.1) from [8] with the initial and final approximations as $\Psi = e^{i(\mathbf{P}_{CI}\mathbf{R}_{CI}' + \mathbf{pr}')/\hbar}\psi_n(\mathbf{r}_1')e^{-i(E_{P_{CI}} + E_p + E_n t/)\hbar}$ and with the "interaction":

$$\hat{V}_{int}(\mathbf{r}_1', \mathbf{r}', \mathbf{R}_{CI}') = V_{pr}(\mathbf{r}') + \delta\hat{H} \quad (A.12)$$

for calculating Born amplitudes.

"Perturbation" (A.11) is a complicated operator whose corrections are not always small and negligible (as we have seen it in analyzing the form-factor expansions). It depends on $\mathbf{R}_{CI}'$, $\mathbf{r}_1'$ and $\mathbf{r}'$ to take into account in a specific way the permanent interaction of atomic constituents. Thus, it is much more preferable to have a problem-free formulation like (A.4) where this taking into account is done exactly – by construction. Then there will not be necessity to sum up some (possibly IR-divergent) series into unavoidable form-factors like $f_n^{n'}(\mathbf{q})$.

## A.3. An IR Analogy with QED

In our formulation (A.8), the interaction potential $V_{pr}(\mathbf{r}')$ (Coulomb field) looks like a "virtual photon" and the bracket term $\delta\hat{H}$ looks like a "coupling to photons". In the first Born approximations in powers of $V_{pr}(\mathbf{r}')$ the nucleus, permanently coupled in the atom, looks decoupled (only elastic amplitude is produced) and the internal atomic degrees of freedom "exist" independently (they are not excited), similarly to what we obtain for a target electron and emitted photons in QED. In fact, it is more than just an analogy. The real electron, as a target, radiates precisely because it is a constituent of a compound system, and when pushed, it excites the target degrees of freedom. However we still have not recognized this physical concept and have not constructed the corresponding mathematical model correctly.

Indeed, originally physicists added, roughly speaking, an interaction including "unknown" self-field to the Hamiltonian by unjustified analogy with the existing, but inexact description – adding a known external potential $V_{pr}(\mathbf{r}')$ to a free Hamiltonian. They immediately obtained wrong self-action effects (corrections to the good phenomenological constants) and then "doctored the wrong numbers" with constant renormalizations (subtractions). The rest was similar to (A.8) with (A9)-(A.12) where summation of the soft contributions to all orders was necessary because "the rest" was an additional, but physically important operator. Later on all these steps were "canonized" as a "gauge way of interaction" furnished with obligatory renormalization and soft diagram summations. Comparison of formulation (A.4) with (A.6) demonstrates that at least the soft part of interaction can be taken into account more effectively, if correctly understood.

Formulation (A.8) does not require renormalization. The standard QED with the renormalized interaction $(\mathcal{L}_{int} + \mathcal{L}_{CT})$ (i.e., with the counter-terms) does not either and it technically gives the results similar to formulation (A.8). Renormalized interaction $(\mathcal{L}_{int} + \mathcal{L}_{CT})$ with $\hat{c}_{\mathbf{k},\lambda}$ and $\hat{c}^{\dagger}_{\mathbf{k},\lambda}$ expressed via combinations of $Q_{\mathbf{k},\lambda}$ and $\partial/\partial Q_{\mathbf{k},\lambda}$ is somewhat similar to (A.12), in my opinion.

### A.4. Mixed variables II – Formulation with Perturbative Corrections to the Initial Constants

The Hydrogen Hamiltonian in the mixed variables, before their rearrangements, has another form:

$$\hat{H}_H = -\frac{\hbar^2}{2M_{Nucl}}\frac{\partial^2}{\partial \mathbf{R}'^2} + \left[-\frac{\hbar^2}{2m_e}\frac{\partial^2}{\partial \mathbf{r}_1'^2} + V_A(\mathbf{r}_1')\right] \\ -\frac{\hbar^2}{2M_{Nucl}}\left(\frac{\partial^2}{\partial \mathbf{r}_1'^2} - 2\frac{\partial}{\partial \mathbf{R}'}\frac{\partial}{\partial \mathbf{r}_1'}\right). \quad (A.13)$$

Note that the masses in the first two terms differ from those of (A.6). Thus, the initial approximation, i.e., the solution to the Schrödinger equations with (A.13) without its round-bracket term $\propto \left(\frac{\partial^2}{\partial \mathbf{r}_1'^2} - 2\frac{\partial}{\partial \mathbf{R}'}\frac{\partial}{\partial \mathbf{r}_1'}\right)$, will also be different from those of (A.6) due to using "wrong" mass values in the good analytical formulas. Here the round-bracket term, considered perturbatively, will take into account not only the missing factor $e^{i\frac{m_e}{M_A}\mathbf{P}\cdot\mathbf{r}_1'/\hbar}$ like in (A.6), but also the masses inexactness in the initial approximations, including the "excitation efficiency" dimensionless constant $m_e/M_A$. Here these corrections to the initial constants are necessary and a great deal of the complicated perturbative corrections here will factually originate from the "mass expansions":

$$M_A^{-1} \approx \frac{1}{M_{Nucl}}(1 - m_e/M_{Nucl} + ...), \quad \mu \approx m_e(1 - m_e/M_{Nucl} + ...)$$

in the results of a simpler formulation (A.6). The "small parameter" here is the ratio $m_e/M_{Nucl}$. Formulation (A.13) is the most awkward one out of all exact formulations since there are many corrections in it, they all are necessary, but there is no physics in those mass expansions/corrections and it is not obvious that in scattering calculations the corresponding corrections to constants can be spotted out and successfully summed up in order to improve the initially wrong masses/constants and simplify the analytical expressions.

(Two other possible forms of the Hydrogen Hamiltonian $\hat{H}_H$, depending on the terms rearrangements, are the following:

$$\hat{H}_H = -\frac{\hbar^2}{2M_A}\frac{\partial^2}{\partial \mathbf{R}'^2} + \left[-\frac{\hbar^2}{2m_e}\frac{\partial^2}{\partial \mathbf{r}_1'^2} + V_A(\mathbf{r}_1')\right] \\ -\frac{\hbar^2}{2M_{Nucl}}\left(\frac{m_e}{M_A}\frac{\partial^2}{\partial \mathbf{R}'^2} + \frac{\partial^2}{\partial \mathbf{r}_1'^2} - 2\frac{\partial}{\partial \mathbf{R}'}\frac{\partial}{\partial \mathbf{r}_1'}\right), \quad (A.14)$$

$$\hat{H}_H - \frac{\hbar^2}{2M_{Nucl}}\frac{\partial^2}{\partial \mathbf{R}'^2} + \left[-\frac{\hbar^2}{2\mu}\frac{\partial^2}{\partial \mathbf{r}_1'^2} + V_A(\mathbf{r}_1')\right] \\ -\frac{\hbar^2}{2M_{Nucl}}\left(-2\frac{\partial}{\partial \mathbf{R}'}\frac{\partial}{\partial \mathbf{r}_1'}\right). \quad (A.15)$$

They too, differ with the mass values in the zeroth-order approximations and with "round-bracket" operators correcting these "wrong starts of perturbation theory" in the frame of the mixed variables formulation.)

### A.5. Further Analogy with QED

Of course, when we understand correctly the coupling physics and our variable change is under our control, we naturally choose the formulation (A.2) for the target spectrum and formulation (A.4) for the scattering problem. The scattering in this case describes the target occupation number evolutions $|\mathbf{P}_A, n\rangle \to |\mathbf{P}'_A, n'\rangle$ due to interaction with a projectile. Our results are simple, comprehensible, and physical. They are easily generalized to the case of a compound projectile with its own spectrum.

However, when we guess the equations in CED and QED, we, roughly speaking, write something like (A.13), i.e., with some additional operator next to $V_{pr}$ or to $V_{ext}$, but with *physical* (measured) numerical constants in the non-perturbed target spectrum $|\mathbf{P}_A, n\rangle$. I.e., we put the values $M_A$ and $\mu$ in the zeroth-order solutions – because these approximations work fine in some important (actually, inclusive) cases:

$$\tilde{\tilde{H}}_H = -\frac{\hbar^2}{2M_A}\frac{\partial^2}{\partial \mathbf{R}'^2} + \left[-\frac{\hbar^2}{2\mu}\frac{\partial^2}{\partial \mathbf{r}_1'^2} + V_A(\mathbf{r}_1')\right] - \frac{\hbar^2}{2M_{Nucl}}\left(\frac{\partial^2}{\partial \mathbf{r}_1'^2} - 2\frac{\partial}{\partial \mathbf{R}'}\frac{\partial}{\partial \mathbf{r}_1'}\right). \quad (A.16)$$

Let me call guess (A.16) a "distorted" (A.13). Factually, it is formulation (A.6), which needs its own round-bracket operator to be correct, not that of (A.13). Thus, the round-bracket term in the "distorted" (A.13) becomes simply wrong. Apart from a "useful" part $\delta\hat{H}$, the round-bracket term in (A.16) contains also a "useless" part whose corrections to the original phenomenological masses/constants, unlike in formulation (A.13), are only harmful and we must get rid of them with "renormalization" (with discarding them in this or that way). Renormalization here is a must and it is just chopping off the unnecessary corrections to good initial constants, just like in QED. In other words, formulation (A.16) with physical constant values in the zeroth-order approximation needs *counter-terms* badly in order to give technically the same results as (A.6). They can even be written explicitly and exactly in our simple case: $L_{CT} = \frac{\hbar^2}{2M_{Nucl}}\left(\frac{\partial^2}{\partial \mathbf{r}_1'^2} - \frac{m_e}{M_A}\frac{\partial^2}{\partial \mathbf{R}'^2}\right)$ or more complicated in terms of the "scattering variables" (A.7) like in (A.11). It is obvious that in such a situation there is no "physics of bare particles" following from some "gauge principles", but there is a bad guess of interaction (see, for example, Introduction in [5]).

And even after perturbative renormalization, the rest of the theory remains awkward since it still needs a selective summation – that of the soft contributions – into reasonable form-factors, like in formulation (A.6). Summation of the soft contributions means, in fact, proceeding from another initial approximation already containing the "expansion parameter" in a non-trivial way - a function instead of a series. After soft contributions summation, the residual series is different from the original one and its convergence properties are different too. There may not be Landau pole, for example, nor Dyson's "proof" of the series being asymptotic, etc. This is how I see the current situation with the standard perturbative QED and that it why I insist on its conceptual and technical reformulation in order to have a problem-free formulation similar to (A.4).

As far as even in the "gauge interaction" the constituents are permanently coupled, my conjecture, therefore, is that a gauge theory furnished with counter-terms $(\mathcal{L}_{int} + \mathcal{L}_{CT})$ is a theory factually formulated in mixed variables a la (A.6), (A.8).

### A.6. "Loops"

I only considered the first Born approximation corresponding to a tree level of QED. A careful reader may wonder why I was mentioning renormalization if I did not consider any loops in my approximation. What about higher-order corrections? Do they include some "loops" or something alike to compare with QED? Let us see.

It is obvious that formulation (A.2) is free from "loop" contributions. Indeed, without any projectile, a free atom solution $|\mathbf{P}_A, n\rangle = |\mathbf{P}_A\rangle |n\rangle$, $E_{\mathbf{P}_A} + E_n = \text{const}$, $\mathbf{P}_A = \text{const}$ (the initial state) stays always the same and does not have any perturbative corrections. However, the zeroth-order solutions in formulation (A.13)-(A.16) get non-trivial perturbative contributions due to the round-bracket terms (kind of "self-action") modifying solely the initial constants even in the absence of any projectile. Both equations – for $\mathbf{R}'$-motion and for $\mathbf{r}_1'$-motion get coupled due to the "round-bracket" operator. Their initially "free" lines (corresponding to solutions $\Phi^{(0)}(\mathbf{R}')$ and $\psi_n^{(0)}(\mathbf{r}_1')$) get kind of self-energy (self-mass) insertions in the first order in the round-bracket operator. One can make sure of it with solving the "perturbed" equations (A.13)-(A.16) with help of their Green's functions and remembering that $E_{\mathbf{P}_A} + E_n = \text{const}$, $\mathbf{P}_A = \text{const}$ [9], [10].

In the scattering calculations the "self-action" operators and the projectile potential mix together which enormously complicates calculations like in QED.

(Formulation (A.6) has corrections to the wave function $\Phi^{(0)}(\mathbf{R}')$ non-reducing to its constant modifications. This formulation is equivalent to formulation (A.4) with interaction $V_{pr}$ expanded "in powers" of $m_e/M_A$ and used as such in all orders of perturbation theory.)

In one-mode model (A.16) the corrections to the initial constants are finite. In QFT there are more normal modes and the resulting perturbative corrections to constants are much bigger (infinite without cutoff). However, finite or infinite, they are unnecessary anyway and are removed under this or that pretext with the corresponding techniques.

A correct formulation must be such that it does not "modify" the phenomenological constants (equation coefficients) in course of calculations, like formulation (A.2), (A.4) giving (1)-(3) in the general case.

# References

[1] Kalitvianski V. (2009). Atom as a "Dressed" Nucleus. *Cent. Eur. J. Phys.*, vol 7, pp. 1-11.; *Preprint* arXiv:0806.2635.

[2] Feynman R. (1964). *The Feynman Lectures on Physics,* vol 2 (Reading, Massachusetts: Addison-Wesley Publishing Company, Inc.), pp. 28-4–28-6.

[3] Kalitvianski V. (2013). A Toy Model of Renormalization and Reformulation. *Int. J. Phys.*, vol 4, pp. 84-93.; *Preprint* arXiv: 1110.3702.

[4] Akhiezer A. I., Berestetskii V. B. (1965). *Quantum Electrodynamics*, vol 2 (New York, USA: Interscience Publishers), p. 413.

[5] Kalitvianski V. (2008). Reformulation Instead of Renormalization, *Preprint* arXiv: 0811.4416.

[6] David Gross. Quantum Field Theory – Past, Present, Future. (2013). *Invited talk at the Conference in Honour of the 90th Birthday of Freeman Dyson,* Nanyang Technological University, Singapore, https://www.youtube.com/watch?v=o1ml3uZGGQE , t=20:55.

[7] Kevin E. Cahill (2011). Renormalization group in continuum quantum field theory and Wilson's views. *Lecture at The Univesity of New Mexico*, https://www.youtube.com/watch?v=3A252xY-a0o

[8] Landau L. D., Lifshitz E. M. (1991). *Quantum Mechanics, Non-relativistic theory*, (New York, USA: Pergamon Press), p. 15.

[9] Leon van Dommelen (2003). The Born series. *Quantum Mechanics for Engineers*. On-line lectures at FAMU-FSU College of Engineering. https://www.eng.fsu.edu/~dommelen/quantum/style_a/nt_bnser.html

[10] Stefan Blügel, (2012) Scattering Theory: Born Series. *Lecture Notes of the 43rd IFF Spring School 2012*. http://juser.fz-juelich.de/record/20885/files/A2_Bluegel.pdf